\documentclass[%
 reprint,
superscriptaddress,
 amsmath,amssymb,
 aps,
]{revtex4-2}

\usepackage{ulem}
\usepackage{graphicx}
\usepackage{dcolumn}
\usepackage{bm}
\usepackage{hyperref}
\hypersetup{
colorlinks=true,
	linkcolor=nrppurple,
	citecolor=nrppurple,
	urlcolor=nrppurple,
}

\newcommand{\df}{\mathrm{d}}

\newcommand{\goes}{\rightarrow}

\newcommand{\w}{\omega}
\newcommand{\W}{\Omega}

\makeatletter \renewcommand\@make@capt@title[2]{%
\@ifx@empty\float@link{\@firstofone}{\expandafter\href\expandafter{\float@link}}%
\sffamily{\textbf{#1}}\@caption@fignum@sep#2 }
\usepackage{placeins} 

\begin{document}
\definecolor{nrppurple}{RGB}{128,0,128}

\preprint{APS/123-QED}

\title{A counterexample to the conjectured Planckian bound on transport}

\author{Nicholas R. Poniatowski}
\email[]{nponiatowski@g.harvard.edu}
\affiliation{Maryland Quantum Materials Center and Department of Physics, University of Maryland, College Park, Maryland 20742, USA}
\affiliation{Department of Physics, Harvard University, Cambridge, MA 02138, USA}
\author{Tarapada Sarkar}
\affiliation{Maryland Quantum Materials Center and Department of Physics, University of Maryland, College Park, Maryland 20742, USA}
\author{Ricardo P.S.M. Lobo}
\affiliation{LPEM, ESPCI Paris, CNRS, PSL University, Paris F-75005, France}
\affiliation{Sorbonne Université, ESPCI Paris, LPEM, Paris F-75005, France}
\author{Sankar Das Sarma}
\affiliation{Condensed Matter Theory Center and Joint Quantum Institute, Department of Physics, University of Maryland, College Park, Maryland 20742, USA}
\author{Richard L. Greene}
\email{rickg@umd.edu}
\affiliation{Maryland Quantum Materials Center and Department of Physics, University of Maryland, College Park, Maryland 20742, USA}

\date{\today}

\begin{abstract}
It has recently been conjectured that the transport relaxation rate in metals is bounded above by the temperature of the system. In this work, we discuss the transport phenomenology of overdoped electron-doped cuprates, which we show constitute an unambiguous counterexample to this putative ``Planckian'' bound, raising serious questions about the efficacy of the bound.
\end{abstract}

\maketitle

The nature of the unconventional metallic state observed in a variety of strongly correlated electronic systems has defied theoretical explanation for the past thirty years, despite being the subject of continuous, intensive research \cite{keimer-review,varma-rmp,our-review,Hussey-review}. The intrigue surrounding these metals is centered around two aspects of their phenomenology: first, they are often claimed to exhibit ``large'' finite-temperature resistivities in excess of the Mott-Ioffe-Regel limit (although the quantitative criterion for this bound is by no means precise \cite{mir-paper,bach}), and are labelled ``bad metals'' \cite{kivelson-badmetal}; and second, they exhibit a linear temperature dependence of the resistivity often down to the lowest experimentally accessible temperatures, and are labelled ``strange metals'' \cite{varma-rmp,our-review}. We emphasize that these two issues, the value of the finite-temperature resistivity and its linear temperature dependence, are completely distinct and should not be conflated: ``bad metallicity'' is an intrinsically finite-temperature phenomenon, whereas the linear-in-$T$ resistivity of the ``strange metal'' pertains to the system's behavior as $T \goes 0$. A system can be bad without being strange and vice versa, or it can be both bad and strange.  Our current work is on a third distinct transport concept much discussed recently, namely, the magnitude of the transport relaxation rate associated with the temperature-dependent resistivity compared with the system temperature itself.

It has recently been found that if one extracts a relaxation timescale $\tau(T)$ from the resistivity by using the Drude formula, $\rho(T) - \rho(0) = \frac{m}{ne^2} \tau^{-1}(T)$, a number of cuprates \cite{legros,admr-planckian} and some other strongly correlated ``strange metals'' \cite{Bruin} satisfy the simple empirical relation
\begin{equation} \label{planckian}
    \tau^{-1}(T) = \alpha \, \frac{k_B T}{\hbar}
\end{equation}
where $\alpha$ is a constant of order unity (note that this subtracts out the $T=0$ contribution to the resistivity from disorder, and focuses entirely on the $T$-dependent part -- this involves a questionable extrapolation of the finite-$T$ resistivity to $T=0$, which we discuss briefly below). In an attempt to explain this striking phenomenological observation, it has been argued that the timescale $\tau(T)$ extracted from the resistivity should be interpreted as either the  electronic momentum relaxation time \cite{zaanen-review} or a many-body equilibration time \cite{hartnoll-review}, and further conjectured that this timescale is subject to a {\it universal} bound $\tau^{-1} \lesssim k_B T/\hbar$. Strongly correlated metals are then hypothesized to saturate this so-called ``Planckian'' limit, which immediately reproduces the empirical relation (\ref{planckian}). We emphasize that this is a conjecture which is loosely rooted in either dimensional analysis arguments (extrapolated from conventional Fermi liquid systems) or in analogy to the Kovtun-Son-Starinets bound \cite{kss} derived for relativistic quantum field theories using techniques from holographic duality \cite{Hartnoll2015,zaanen-review}, although no convincing arguments exist for why such a field theoretic bound should be related to the electrical transport properties of solid state materials. A key feature of the conjectured Planckian bound is that it is purported to be universal: that is, it is a general principle of nature which should apply to {\it all} physical systems at all temperatures. Purely physically, a bound on a relaxation rate imposed by the system temperature sounds reasonable based on the vague idea that relaxation or dissipation should not exceed the temperature, but to date no proof of this exists.

We emphasize that the existence of a Planckian bound on transport is an issue that is distinct from the broader mysteries of linear-in-$T$ resistivity, and Eq. (\ref{planckian}), as posed, applies independent of the detailed functional form of $\rho(T)$ or $\tau(T)$. That is, the existence or non-existence of such a bound is unrelated to the validity (or not) of the quasiparticle picture, let alone the underlying physical scattering mechanism responsible for the linear-in-$T$ resistivity. As long as a resistivity can be measured as a function of temperature enabling the extraction of the effective transport relaxation time from the Drude formula, the putative Planckian bound should apply. Further, there is no consensus on the theoretical ``meaning'' of the Planckian bound (including on what timescale is actually subject to the bound), and several schools of thought continue to develop \cite{zaanen-review,hartnoll-review,Hartnoll2015}. In light of this, we take a conservative approach, eschewing such theoretical questions, and limiting our discussion to the operational, {\it experimental} question: can the relaxation rate $\tau^{-1}$ extracted from the resistivity using the Drude formula exceed the system temperature in a strongly correlated system?

Amusingly, it has also been realized that even many conventional ``good'' metals such as Cu, Ag, and Au are ``Planckian'' \cite{Bruin}. The reason is well understood, originating from the Migdal-Eliashberg theory of electron-phonon interactions, wherein the electronic relaxation rate is given by $\tau^{-1} = 2\pi \lambda \,k_B T/\hbar$ above the Debye temperature (more precisely the Bloch-Gruneisen temperature), with $\lambda$ the dimensionless electron-phonon coupling constant \cite{ALLEN19831,Grimvall_1976}. In general, $\lambda \sim 0.1$ for weakly coupled metals such as Cu, Ag, or Au, leading to a ``Planckian'' behavior where $2\pi \lambda \sim 1$, agreeing with Eq. (\ref{planckian}). In fact, this behavior was comprehensively studied in the 1960s in a series of now-classic papers \cite{allen-lambdatr}, and the fact that simple metals trivially obey the Planckian bound with no deep significance is well known. Further, one can consider metals with strong electron-phonon coupling such as Nb or Pb, where $\lambda \gtrsim 1$ and the momentum relaxation rate is of the form of Eq. (\ref{planckian}) with $\alpha \approx 6 - 7.5$, far in excess of the Planckian limit of $\alpha \approx 1$. This of course is neither mysterious nor deep as it arises from electron-acoustic phonon scattering occurring in a regime where the phonon thermal occupancy follows the classical equipartition (linear-in-$T$) behavior. This has been emphasized recently in the theoretical literature \cite{sds-magic,Sadovskii_2021,sds-dilute}. In particular, strange metal and Planckian behavior in twisted bilayer graphene with high resitivity \cite{mit-planckian,graphene-linear} has recently been explained based on electron-phonon interactions \cite{sds-magic}. 

In principle, this phonon-induced large magnitude linear-in-$T$ resistivity can increase indefinitely (and thus lead to arbitrarily large violations of the Planckian bound), until one approaches the Fermi temperature. For low density (small $k_F$) systems such as the cuprates, this linear-in-$T$ behavior can persist down to low temperatures since the lower cutoff scale (the Bloch-Gruneisen temperature) can be small in dilute systems \cite{sds-magic,sds-dilute}. Moreover, we note that conventional electron-phonon systems at high temperature are ``non-quasiparticle'' metals according to the usual criteria $\tau_{\text{qp}}^{-1} \sim T$, and yet are still described extremely effectively using semi-classical transport theory \cite{sds-trivial-nfl}. The key, of course, is that the quasi-particle is defined close to the Fermi surface at low temperatures, and not by the behavior of a relaxation rate at `high' temperatures \cite{sds-trivial-nfl}. 

It has recently been argued \cite{hartnoll-review} that high-temperature electron-phonon scattering is immune from the Planckian bound in that the scattering is quasi-elastic. However, this requires a suitable choice of the timescale which is subject to the Planckian bound (an issue which there is currently no clear consensus on). One might also question the practical utility of conjecturing a bound on an experimentally inaccessible quantity. The serious conceptual problem in this line of reasoning is that the experimental resistivity simply produces a relaxation rate through the Drude formula, and there is no way to know a priori or a posteriori whether this  transport relaxation rate is elastic or inelastic, or if it arises from scattering by phonons or some other excitations.  Thus, while theoretically it makes perfect sense to impose a bound only on an inelastic relaxation rate arising from non-phonon mechanisms, this is meaningless in understanding a specific system (e.g. cuprates) manifesting (or not) a Planckian behavior since by definition we do not know the mechanism causing the resistive scattering. Nonetheless, at a phenomenological level, the electron-phonon system unambiguously demonstrates that one can find transport relaxation rates in excess of the system temperature without needing to invoke any exotic physics.

It is also important to emphasize that at $T=0$ or more generally at low temperatures, any finite resistivity, by definition, violates the Planckian bound since the corresponding relaxation rate, being finite, trivially exceeds the temperature by an arbitrary factor.  This is the reason that the definition of the Planckian bound often involves a subtraction of the $T=0$ resistivity, but such a $T=0$ extrapolated subtraction, even if it can be done accurately (which is always a question),  assumes that the resistive scattering mechanism at $T=0$ (i.e. disorder scattering) is temperature-independent, which is often not the case since disorder may have effective temperature dependence arising from weak localization, interaction, screening, and annealing effects. In fact, it is well-known that the low-temperature resistivity of a specific sample often changes as the sample is thermally cycled because of impurity annealing effects.

\begin{figure}
    \centering
    \includegraphics[width=90mm]{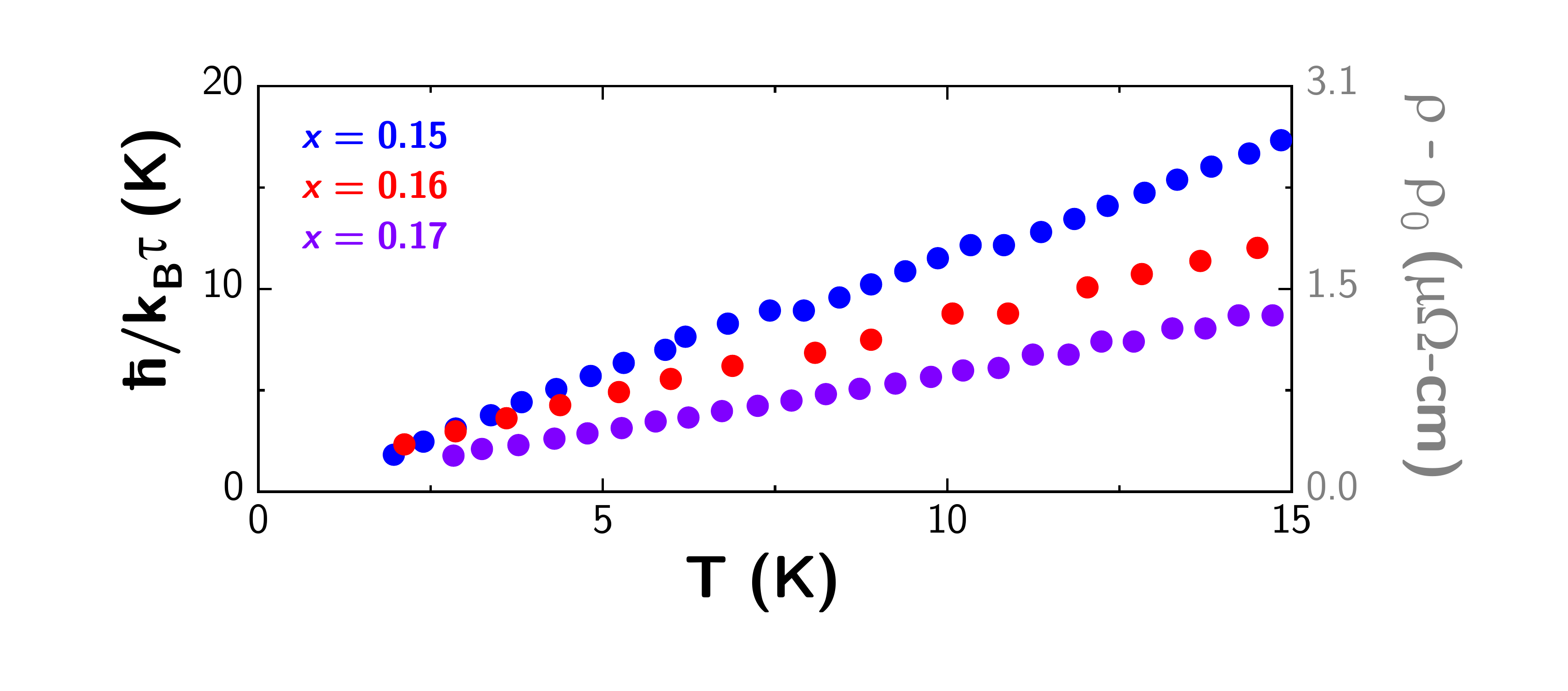}
    \caption{Low temperature linear-in-$T$ resistivity of LCCO for three different dopings. The resistivity curves were taken in a 8 T magnetic field for the $x = 0.15$ sample, 6.5 T for the $x = 0.16$ sample, and 4 T for the $x = 0.17$ sample. The residual resistivity $\rho_0 = \rho(T \goes0)$ is subtracted from each curve, and the relaxation rate $1/ \tau = (ne^2/m_\star) (\rho(T)-\rho_0)$ is extracted using the parameters $n = 9\times 10^{27}~\mathrm{m}^{-3}$ and $m_\star = 3m_{\text{e}}$ (with $m_{\text{e}}$ the electron mass) used in Ref. \cite{legros}. }
    \label{low-T-fig}
\end{figure}

In this work, we consider the electron-doped family of the cuprate high-temperature superconductors which includes the compounds La$_{2-x}$Ce$_x$CuO$_4$ (LCCO) and Pr$_{2-x}$Ce$_x$CuO$_4$ (PCCO)\cite{rick-rmp}. For dopings above the Fermi surface reconstruction doping ($x_{\text{FSR}} = 0.14$ for LCCO \cite{tara-prb} and $x_{\text{FSR} }= 0.17$ for PCCO \cite{yoram-prl}), the resistivities of both LCCO \cite{kui} and PCCO \cite{pcco-linear-t} are linear down to mK temperatures when superconductivity is suppressed with a modest magnetic field (less than 9 T for LCCO, and 12 T for PCCO). To illustrate this, we plot the low-temperature resistivity for three dopings of LCCO in Fig. \ref{low-T-fig} (the strangeness of linear-in-$T$ resistivity persisting over a finite range of dopings is tangential to the subject of this work, but is extensively discussed in Ref. \onlinecite{our-review}).

Further, it has recently been observed \cite{legros} that the relaxation time $\tau(T)$ extracted from the resistivity for each of these compounds can be fit to the Planckian form (\ref{planckian}) with $\alpha = 1.2 \pm 0.3$ for $x=0.15$ LCCO and $\alpha = 0.8 \pm 0.2$ for $x = 0.17$ PCCO. Following the analysis of Ref. \onlinecite{legros}, we convert the resistivity of the three LCCO samples in Fig. \ref{low-T-fig} to the relaxation time $\tau$ using the Drude equation with the parameters $n = 9\times 10^{27}~\mathrm{m}^{-3}$ and $m_\star = 3m_{\text{e}}$ where $m_{\text{e}}$ is the electron mass. In agreement with Ref. \onlinecite{legros}, we find that the low-temperature resistivity is consistent with the Planckian bound Eq. (\ref{planckian}), with $\alpha \approx 1$. Although there is some controversy in the literature \cite{varma-rmp,Sadovskii_2021} surrounding the analysis of Ref. \onlinecite{legros}, and the extraction of timescales from the Drude equation in general, we take this procedure to at least be representative of the typical analyses underlying claims of ``Planckian'' behavior. The choice of different parameters (i.e. $n$ or $m_\star$) will not qualitatively affect any of the conclusions reached below.

\begin{figure}
    \centering
    \includegraphics[width=80mm]{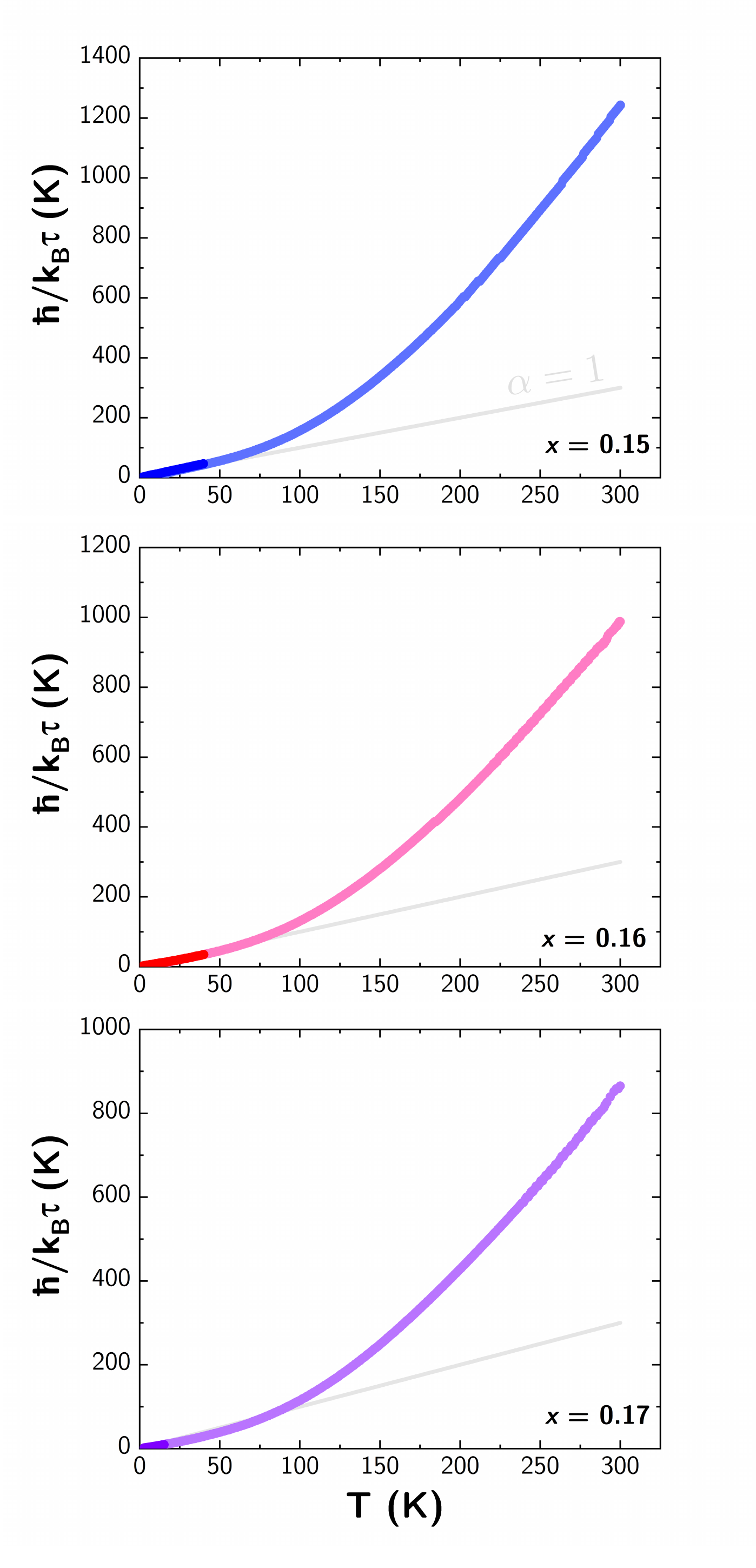}
    \caption{The relaxation rate $\tau^{-1}$ (in temperature units) as a function of temperature for La$_{2-x}$Ce$_x$CuO$_4$ for $x = 0.15,~0.16,~0.17$. The dark colored curves in each plot are the low-temperature relaxation rates plotted previously in Fig. \ref{low-T-fig} and the lighter colored curves are the relaxation rates measured up to room temperature in zero magnetic field. The light grey curves in each plot indicate the Planckian limit $\hbar/k_B\tau = T$. }
    \label{the-figure}
\end{figure}

Crucially, Ref. \onlinecite{legros} only considered the response of the system for $T < 30$ K, while it is well-known that at higher temperatures, above $\sim$ 50 K, the resistivity of electron-doped cuprates crosses over to a $T^2$-power law, which persists up to $>$ 400 K \cite{fl-sm,our-review}. To give a more complete picture, in Fig. \ref{the-figure} we present the full temperature dependence of the relaxation rate $\tau^{-1}$ associated with the resistivity (extracted following the same procedure as described above), from 2 K to 300 K for the three LCCO dopings shown previously in Fig. \ref{low-T-fig}.

Doing so, we see that $\tau^{-1}$ for LCCO drastically exceeds the putative Planckian bound, with $\tau^{-1} \gg k_B T/\hbar$ once the $T^2$ resistivity sets in (we emphasize again that the Planckian bound does not stipulate any specific $T$-dependence of  $\tau(T)$ or $\rho(T)$, asserting its generality). Further, the temperature-dependence of $\tau^{-1}$ is super-linear, parametrically violating the Planckian bound at high temperatures. This simple observation unambiguously establishes that one can find a relaxation rate, directly extracted from the Drude formula, that is well in excess of the system temperature. In other words, this serves as an explicit counterexample to the Planckian bound as it is typically applied to real materials at a phenomenological level.

We may now go beyond the level of analysis typically employed in claims of Planckian behavior, and consider the possible origins of the observed temperature dependence of the resistivity. Appealing to the Drude equation, the temperature-dependent resistivity has two contributions, each of which is possibly temperature-dependent. First, one has the relaxation rate $\tau^{-1}$ that is conjectured to be subject to the Planckian bound. In most works claiming Planckian physics, it is implicitly assumed that all of the temperature dependence is carried by $\tau^{-1}$. However, the Drude weight $ne^2/m_\star$ can, in principle, also be temperature dependent. That is, the temperature dependence of the resistivity need not necessarily be the temperature dependence of $\tau^{-1}$.

\begin{figure}
    \centering
    \includegraphics[width=85mm]{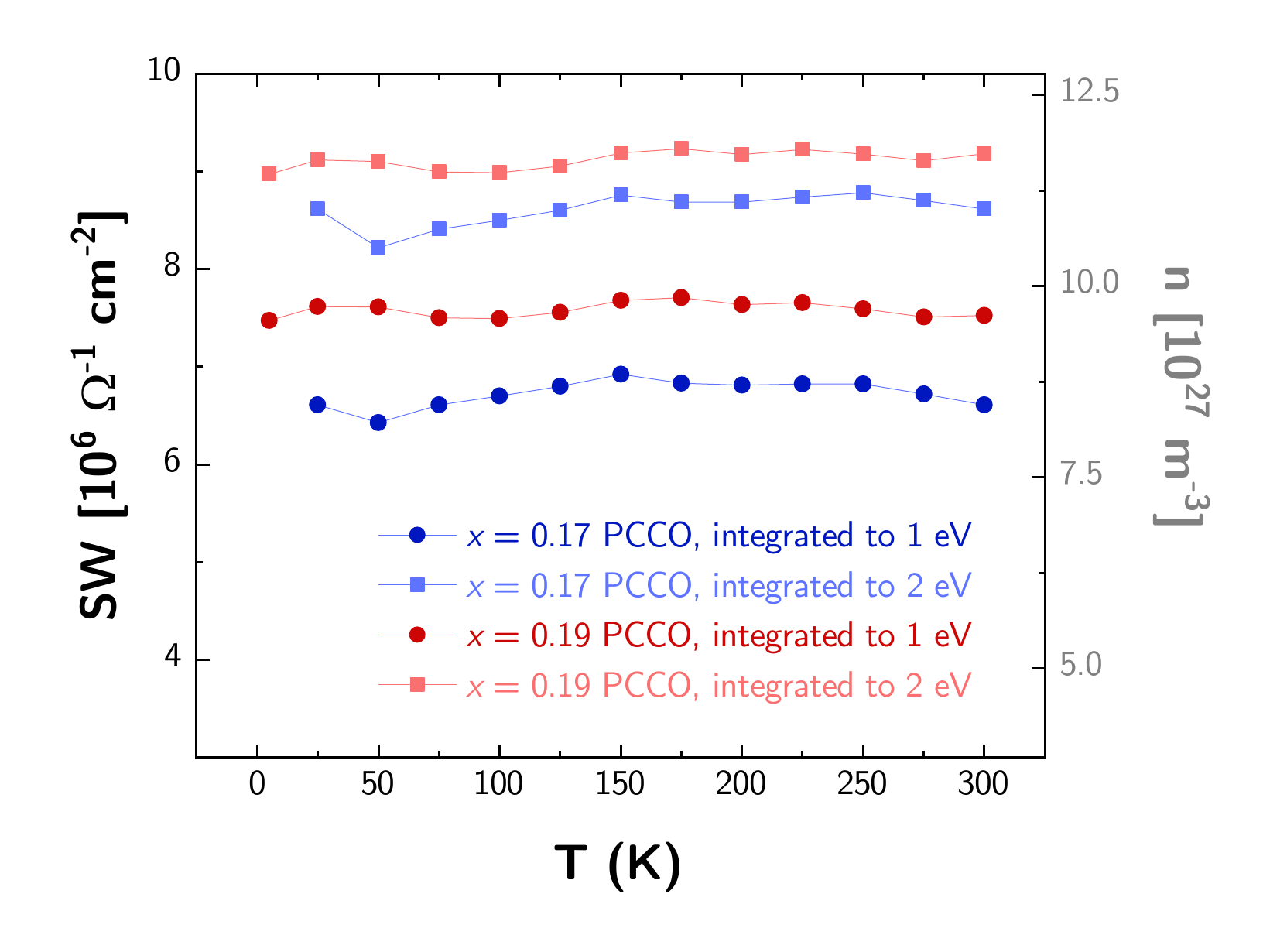}
    \caption{Temperature-dependent optical conductivity spectral
weight (SW) for $x = 0.17$ and $x = 0.19$ PCCO samples with upper cut-offs of 1 eV and 2 eV, representative of the conduction band width. On the right axis, the spectral weight is converted to the associated carrier density through $n = ((2\pi c)^2 \varepsilon_0 Z_0 m_\star)/(\pi^2 e^2)\int \df \w \; \sigma_1(\w)$ where $c$ is the speed of light, $\varepsilon_0$ is the vacuum permitivitty, $Z_0$ is the vacuum impedance, and the spectral weight is in units of $\W^{-1} \, \mathrm{cm}^{-2}$.  Assuming that $m_\star = 3 m_{\text{e}}$, the carrier density directly inferred from the integrated spectral weight is in good agreement with the value $n = 9 \times 10^{27} \; \mathrm{m}^{-3}$ for LCCO used in our analysis above. Note that data below 20 K for the $x=0.17$ sample is not shown as superconductivity depletes the finite frequency spectral weight.}
    \label{sumrule-fig}
\end{figure}

The Drude weight $ne^2/m_\star$ can be directly inferred from the integrated spectral weight of the optical conductivity. The optical conductivity $f$-sum rule states that the integral of
the real part of the optical conductivity from zero to infinity is
proportional to the ratio $n_{\text{e}} / m_{\text{e}}$, where $n_{\text{e}}$ is the total density of
electrons in the material and $m_{\text{e}}$ is the bare electronic mass. In a conductor, cutting the integral off at some finite value comparable to
the band width gives a good estimation of the value of $n / m^\star$,
\begin{equation}
    \int_0^\Lambda \df \omega  \; \sigma_1(\omega) = \frac{\pi}{2} \frac{n e^2}{m_\star} \, ,
\end{equation}
where $n$ is the carrier density and $\Lambda$ is the upper frequency cut-off. Figure \ref{sumrule-fig} shows the
optical conductivity spectral weight with upper cut-offs of 1 eV and 2 eV
for both $x = 0.17$ \cite{Zimmers_2005} and $0.19$ (this work) PCCO. Both of these samples are doped above the Fermi surface reconstruction doping for PCCO $(x_{\text{FSR}} = 0.17$), and are comparable to the overdoped $x = 0.15-0.17$ LCCO samples discussed above. The plasma edge in both materials lies close to 1 eV and hence these cut-off values are good approximations for the conduction band width. Crucially, Fig. \ref{sumrule-fig} shows
that the $n / m^\star$ ratio is essentially temperature independent for
both chosen cut-offs. In light of this, we can confidently attribute the temperature dependence of $\rho(T)$ fully to that of the relaxation rate $\tau^{-1}$, and consequently be certain that the Planckian bound is indeed strongly violated in our system.

One may also attempt to explain the gross violation of the Planckian bound in electron-doped cuprates by conjecturing that the super-Planckian scattering is quasi-elastic and thus not subject to the Planckian bound (as has been argued \cite{hartnoll-review} to explain the violation of the Planckian bound in conventional electron-phonon systems, discussed above). However, this conjecture is wholly unsupported by all existing experimental evidence. The high-temperature $\rho(T) \sim T^2$ behavior is incompatible with a conventional phonon-based description which should produce a linear-in-$T$ resistivity in this temperature regime (as in normal metals) \cite{fl-sm}, and although the scattering mechanism underlying this unconventional behavior is unknown, it is likely of electronic origin, and hence should be subject to the Planckian bound. This is further reinforced by the fact that inelastic electron-electron scattering leads to a $T^2$ resistivity in many correlated materials (e.g. heavy fermions, pnictides, transition metal oxides, and organic metals). Thus, there is no experimentally justifiable reason to expect that the electron-doped cuprates are exempt from the Planckian bound. Moreover, the Planckian bound is a rather empty concept if one must first decide a priori whether a relaxation rate is elastic or inelastic before imposing the bound since the measurement of resistivity simply produces a momentum relaxation rate, which is allowed to be elastic or inelastic. Imposing the severe constraint that the bound is only relevant when the scattering is inelastic makes the bound a semantic exercise, particularly since many normal metals obey the bound trivially by virtue of electron-phonon scattering in the equipartition regime, as discussed above. This is also true if the bound is constrained to apply only near quantum criticality since one has no guarantee that the measured resistivity arises from inelastic quantum critical scattering. After all, the bound seems to be obeyed by weakly coupled metals, but not by strongly coupled metals, with quantum criticality playing no role in either. 

 Indeed, if a metallic resistivity is linear in temperature down to arbitrarily low temperatures, it is an extremely interesting and exotic situation, quite independent of any Planckian implications, since such a linearity at very low temperatures implies a possible manifest non-Fermi liquid behavior if the underlying mechanism is inelastic electron-electron interactions, and not elastic impurity and/or phonon scattering.  But our Planckian-violating resistivity goes as $T^2$ at high temperatures, and, as discussed above, the $T^2$ behavior is commonly associated with Fermi liquids.  Phonon or impurity scattering is not known to produce a $T^2$ resistivity in any regime.

To summarize, we have shown that the electron-doped cuprates provide a clear and strong counterexample to the conjectured Planckian bound on electronic transport. In so doing, we have shown that the Planckian bound can be violated in a strongly correlated cuprate system, just as it is violated in conventional electron-phonon systems and dilute semiconductors \cite{sto-prx}. Our work shows that the Planckian bound is not universal and that invoking it is of limited practical utility. 


\begin{acknowledgments}
The experimental work was supported by the NSF under Grant No. DMR-2002658 and the Maryland Quantum Materials Center. N.R.P. is supported by the Army Research Office through an NDSEG fellowship. S.D.S. is supported by the Laboratory for Physical Sciences. 
\end{acknowledgments}

\bibliography{apssamp}

\begin{thebibliography}{31}%
\makeatletter
\providecommand \@ifxundefined [1]{%
 \@ifx{#1\undefined}
}%
\providecommand \@ifnum [1]{%
 \ifnum #1\expandafter \@firstoftwo
 \else \expandafter \@secondoftwo
 \fi
}%
\providecommand \@ifx [1]{%
 \ifx #1\expandafter \@firstoftwo
 \else \expandafter \@secondoftwo
 \fi
}%
\providecommand \natexlab [1]{#1}%
\providecommand \enquote  [1]{``#1''}%
\providecommand \bibnamefont  [1]{#1}%
\providecommand \bibfnamefont [1]{#1}%
\providecommand \citenamefont [1]{#1}%
\providecommand \href@noop [0]{\@secondoftwo}%
\providecommand \href [0]{\begingroup \@sanitize@url \@href}%
\providecommand \@href[1]{\@@startlink{#1}\@@href}%
\providecommand \@@href[1]{\endgroup#1\@@endlink}%
\providecommand \@sanitize@url [0]{\catcode `\\12\catcode `\$12\catcode
  `\&12\catcode `\#12\catcode `\^12\catcode `\_12\catcode `\%12\relax}%
\providecommand \@@startlink[1]{}%
\providecommand \@@endlink[0]{}%
\providecommand \url  [0]{\begingroup\@sanitize@url \@url }%
\providecommand \@url [1]{\endgroup\@href {#1}{\urlprefix }}%
\providecommand \urlprefix  [0]{URL }%
\providecommand \Eprint [0]{\href }%
\providecommand \doibase [0]{https://doi.org/}%
\providecommand \selectlanguage [0]{\@gobble}%
\providecommand \bibinfo  [0]{\@secondoftwo}%
\providecommand \bibfield  [0]{\@secondoftwo}%
\providecommand \translation [1]{[#1]}%
\providecommand \BibitemOpen [0]{}%
\providecommand \bibitemStop [0]{}%
\providecommand \bibitemNoStop [0]{.\EOS\space}%
\providecommand \EOS [0]{\spacefactor3000\relax}%
\providecommand \BibitemShut  [1]{\csname bibitem#1\endcsname}%
\let\auto@bib@innerbib\@empty
\bibitem [{\citenamefont {Keimer}\ \emph {et~al.}(2015)\citenamefont {Keimer},
  \citenamefont {Kivelson}, \citenamefont {Norman}, \citenamefont {Uchida},\
  and\ \citenamefont {Zaanen}}]{keimer-review}%
  \BibitemOpen
  \bibfield  {author} {\bibinfo {author} {\bibfnamefont {B.}~\bibnamefont
  {Keimer}}, \bibinfo {author} {\bibfnamefont {S.~A.}\ \bibnamefont
  {Kivelson}}, \bibinfo {author} {\bibfnamefont {M.~R.}\ \bibnamefont
  {Norman}}, \bibinfo {author} {\bibfnamefont {S.}~\bibnamefont {Uchida}},\
  and\ \bibinfo {author} {\bibfnamefont {J.}~\bibnamefont {Zaanen}},\
  }\bibfield  {title} {\bibinfo {title} {From quantum matter to
  high-temperature superconductivity in copper oxides},\ }\bibfield  {journal}
  {\bibinfo  {journal} {Nature (London)}\ }\textbf {\bibinfo {volume} {518}},\
  \href {https://doi.org/10.1038/nature14165} {10.1038/nature14165} (\bibinfo
  {year} {2015})\BibitemShut {NoStop}%
\bibitem [{\citenamefont {Varma}(2020)}]{varma-rmp}%
  \BibitemOpen
  \bibfield  {author} {\bibinfo {author} {\bibfnamefont {C.~M.}\ \bibnamefont
  {Varma}},\ }\bibfield  {title} {\bibinfo {title} {Colloquium: Linear in
  temperature resistivity and associated mysteries including high temperature
  superconductivity},\ }\href {https://doi.org/10.1103/RevModPhys.92.031001}
  {\bibfield  {journal} {\bibinfo  {journal} {Rev. Mod. Phys.}\ }\textbf
  {\bibinfo {volume} {92}},\ \bibinfo {pages} {031001} (\bibinfo {year}
  {2020})}\BibitemShut {NoStop}%
\bibitem [{\citenamefont {Greene}\ \emph {et~al.}(2020)\citenamefont {Greene},
  \citenamefont {Mandal}, \citenamefont {Poniatowski},\ and\ \citenamefont
  {Sarkar}}]{our-review}%
  \BibitemOpen
  \bibfield  {author} {\bibinfo {author} {\bibfnamefont {R.~L.}\ \bibnamefont
  {Greene}}, \bibinfo {author} {\bibfnamefont {P.~R.}\ \bibnamefont {Mandal}},
  \bibinfo {author} {\bibfnamefont {N.~R.}\ \bibnamefont {Poniatowski}},\ and\
  \bibinfo {author} {\bibfnamefont {T.}~\bibnamefont {Sarkar}},\ }\bibfield
  {title} {\bibinfo {title} {The strange metal state of the electron-doped
  cuprates},\ }\href {https://doi.org/10.1146/annurev-conmatphys-031119-050558}
  {\bibfield  {journal} {\bibinfo  {journal} {Annual Review of Condensed Matter
  Physics}\ }\textbf {\bibinfo {volume} {11}},\ \bibinfo {pages} {213}
  (\bibinfo {year} {2020})},\ \Eprint
  {https://arxiv.org/abs/https://doi.org/10.1146/annurev-conmatphys-031119-050558}
  {https://doi.org/10.1146/annurev-conmatphys-031119-050558} \BibitemShut
  {NoStop}%
\bibitem [{\citenamefont {Hussey}(2008)}]{Hussey-review}%
  \BibitemOpen
  \bibfield  {author} {\bibinfo {author} {\bibfnamefont {N.~E.}\ \bibnamefont
  {Hussey}},\ }\bibfield  {title} {\bibinfo {title} {Phenomenology of the
  normal state in-plane transport properties of high-{T$_c$} cuprates},\ }\href
  {https://doi.org/10.1088/0953-8984/20/12/123201} {\bibfield  {journal}
  {\bibinfo  {journal} {Journal of Physics: Condensed Matter}\ }\textbf
  {\bibinfo {volume} {20}},\ \bibinfo {pages} {123201} (\bibinfo {year}
  {2008})}\BibitemShut {NoStop}%
\bibitem [{\citenamefont {Poniatowski}\ \emph {et~al.}(2021)\citenamefont
  {Poniatowski}, \citenamefont {Sarkar}, \citenamefont {Das~Sarma},\ and\
  \citenamefont {Greene}}]{mir-paper}%
  \BibitemOpen
  \bibfield  {author} {\bibinfo {author} {\bibfnamefont {N.~R.}\ \bibnamefont
  {Poniatowski}}, \bibinfo {author} {\bibfnamefont {T.}~\bibnamefont {Sarkar}},
  \bibinfo {author} {\bibfnamefont {S.}~\bibnamefont {Das~Sarma}},\ and\
  \bibinfo {author} {\bibfnamefont {R.~L.}\ \bibnamefont {Greene}},\ }\bibfield
   {title} {\bibinfo {title} {Resistivity saturation in an electron-doped
  cuprate},\ }\href {https://doi.org/10.1103/PhysRevB.103.L020501} {\bibfield
  {journal} {\bibinfo  {journal} {Phys. Rev. B}\ }\textbf {\bibinfo {volume}
  {103}},\ \bibinfo {pages} {L020501} (\bibinfo {year} {2021})}\BibitemShut
  {NoStop}%
\bibitem [{\citenamefont {Bach}\ \emph {et~al.}(2011)\citenamefont {Bach},
  \citenamefont {Saha}, \citenamefont {Kirshenbaum}, \citenamefont {Paglione},\
  and\ \citenamefont {Greene}}]{bach}%
  \BibitemOpen
  \bibfield  {author} {\bibinfo {author} {\bibfnamefont {P.~L.}\ \bibnamefont
  {Bach}}, \bibinfo {author} {\bibfnamefont {S.~R.}\ \bibnamefont {Saha}},
  \bibinfo {author} {\bibfnamefont {K.}~\bibnamefont {Kirshenbaum}}, \bibinfo
  {author} {\bibfnamefont {J.}~\bibnamefont {Paglione}},\ and\ \bibinfo
  {author} {\bibfnamefont {R.~L.}\ \bibnamefont {Greene}},\ }\bibfield  {title}
  {\bibinfo {title} {High-temperature resistivity in the iron pnictides and the
  electron-doped cuprates},\ }\href
  {https://doi.org/10.1103/PhysRevB.83.212506} {\bibfield  {journal} {\bibinfo
  {journal} {Phys. Rev. B}\ }\textbf {\bibinfo {volume} {83}},\ \bibinfo
  {pages} {212506} (\bibinfo {year} {2011})}\BibitemShut {NoStop}%
\bibitem [{\citenamefont {Emery}\ and\ \citenamefont
  {Kivelson}(1995)}]{kivelson-badmetal}%
  \BibitemOpen
  \bibfield  {author} {\bibinfo {author} {\bibfnamefont {V.~J.}\ \bibnamefont
  {Emery}}\ and\ \bibinfo {author} {\bibfnamefont {S.~A.}\ \bibnamefont
  {Kivelson}},\ }\bibfield  {title} {\bibinfo {title} {Superconductivity in bad
  metals},\ }\href {https://doi.org/10.1103/PhysRevLett.74.3253} {\bibfield
  {journal} {\bibinfo  {journal} {Phys. Rev. Lett.}\ }\textbf {\bibinfo
  {volume} {74}},\ \bibinfo {pages} {3253} (\bibinfo {year}
  {1995})}\BibitemShut {NoStop}%
\bibitem [{\citenamefont {Legros}\ \emph {et~al.}(2019)\citenamefont {Legros},
  \citenamefont {Benhabib}, \citenamefont {Tabis}, \citenamefont
  {Lalibert{\'e}}, \citenamefont {Dion}, \citenamefont {Lizaire}, \citenamefont
  {Vignolle}, \citenamefont {Vignolles}, \citenamefont {Raffy}, \citenamefont
  {Li}, \citenamefont {Auban-Senzier}, \citenamefont {Doiron-Leyraud},
  \citenamefont {Fournier}, \citenamefont {Colson}, \citenamefont {Taillefer},\
  and\ \citenamefont {Proust}}]{legros}%
  \BibitemOpen
  \bibfield  {author} {\bibinfo {author} {\bibfnamefont {A.}~\bibnamefont
  {Legros}}, \bibinfo {author} {\bibfnamefont {S.}~\bibnamefont {Benhabib}},
  \bibinfo {author} {\bibfnamefont {W.}~\bibnamefont {Tabis}}, \bibinfo
  {author} {\bibfnamefont {F.}~\bibnamefont {Lalibert{\'e}}}, \bibinfo {author}
  {\bibfnamefont {M.}~\bibnamefont {Dion}}, \bibinfo {author} {\bibfnamefont
  {M.}~\bibnamefont {Lizaire}}, \bibinfo {author} {\bibfnamefont
  {B.}~\bibnamefont {Vignolle}}, \bibinfo {author} {\bibfnamefont
  {D.}~\bibnamefont {Vignolles}}, \bibinfo {author} {\bibfnamefont
  {H.}~\bibnamefont {Raffy}}, \bibinfo {author} {\bibfnamefont {Z.~Z.}\
  \bibnamefont {Li}}, \bibinfo {author} {\bibfnamefont {P.}~\bibnamefont
  {Auban-Senzier}}, \bibinfo {author} {\bibfnamefont {N.}~\bibnamefont
  {Doiron-Leyraud}}, \bibinfo {author} {\bibfnamefont {P.}~\bibnamefont
  {Fournier}}, \bibinfo {author} {\bibfnamefont {D.}~\bibnamefont {Colson}},
  \bibinfo {author} {\bibfnamefont {L.}~\bibnamefont {Taillefer}},\ and\
  \bibinfo {author} {\bibfnamefont {C.}~\bibnamefont {Proust}},\ }\bibfield
  {title} {\bibinfo {title} {Universal {T}-linear resistivity and {P}lanckian
  dissipation in overdoped cuprates},\ }\href
  {https://doi.org/10.1038/s41567-018-0334-2} {\bibfield  {journal} {\bibinfo
  {journal} {Nature Physics}\ }\textbf {\bibinfo {volume} {15}},\ \bibinfo
  {pages} {142} (\bibinfo {year} {2019})}\BibitemShut {NoStop}%
\bibitem [{\citenamefont {Grissonnanche}\ \emph {et~al.}(2021)\citenamefont
  {Grissonnanche}, \citenamefont {Fang}, \citenamefont {Legros}, \citenamefont
  {Verret}, \citenamefont {Lalibert{\'e}}, \citenamefont {Collignon},
  \citenamefont {Zhou}, \citenamefont {Graf}, \citenamefont {Goddard},
  \citenamefont {Taillefer},\ and\ \citenamefont {Ramshaw}}]{admr-planckian}%
  \BibitemOpen
  \bibfield  {author} {\bibinfo {author} {\bibfnamefont {G.}~\bibnamefont
  {Grissonnanche}}, \bibinfo {author} {\bibfnamefont {Y.}~\bibnamefont {Fang}},
  \bibinfo {author} {\bibfnamefont {A.}~\bibnamefont {Legros}}, \bibinfo
  {author} {\bibfnamefont {S.}~\bibnamefont {Verret}}, \bibinfo {author}
  {\bibfnamefont {F.}~\bibnamefont {Lalibert{\'e}}}, \bibinfo {author}
  {\bibfnamefont {C.}~\bibnamefont {Collignon}}, \bibinfo {author}
  {\bibfnamefont {J.}~\bibnamefont {Zhou}}, \bibinfo {author} {\bibfnamefont
  {D.}~\bibnamefont {Graf}}, \bibinfo {author} {\bibfnamefont {P.~A.}\
  \bibnamefont {Goddard}}, \bibinfo {author} {\bibfnamefont {L.}~\bibnamefont
  {Taillefer}},\ and\ \bibinfo {author} {\bibfnamefont {B.~J.}\ \bibnamefont
  {Ramshaw}},\ }\bibfield  {title} {\bibinfo {title} {Linear-in temperature
  resistivity from an isotropic {P}lanckian scattering rate},\ }\href
  {https://doi.org/10.1038/s41586-021-03697-8} {\bibfield  {journal} {\bibinfo
  {journal} {Nature}\ }\textbf {\bibinfo {volume} {595}},\ \bibinfo {pages}
  {667} (\bibinfo {year} {2021})}\BibitemShut {NoStop}%
\bibitem [{\citenamefont {Bruin}\ \emph {et~al.}(2013)\citenamefont {Bruin},
  \citenamefont {Sakai}, \citenamefont {Perry},\ and\ \citenamefont
  {Mackenzie}}]{Bruin}%
  \BibitemOpen
  \bibfield  {author} {\bibinfo {author} {\bibfnamefont {J.~A.~N.}\
  \bibnamefont {Bruin}}, \bibinfo {author} {\bibfnamefont {H.}~\bibnamefont
  {Sakai}}, \bibinfo {author} {\bibfnamefont {R.~S.}\ \bibnamefont {Perry}},\
  and\ \bibinfo {author} {\bibfnamefont {A.~P.}\ \bibnamefont {Mackenzie}},\
  }\bibfield  {title} {\bibinfo {title} {Similarity of scattering rates in
  metals showing {T}-linear resistivity},\ }\href
  {https://doi.org/10.1126/science.1227612} {\bibfield  {journal} {\bibinfo
  {journal} {Science}\ }\textbf {\bibinfo {volume} {339}},\ \bibinfo {pages}
  {804} (\bibinfo {year} {2013})},\ \Eprint
  {https://arxiv.org/abs/https://science.sciencemag.org/content/339/6121/804.full.pdf}
  {https://science.sciencemag.org/content/339/6121/804.full.pdf} \BibitemShut
  {NoStop}%
\bibitem [{\citenamefont {Zaanen}(2019)}]{zaanen-review}%
  \BibitemOpen
  \bibfield  {author} {\bibinfo {author} {\bibfnamefont {J.}~\bibnamefont
  {Zaanen}},\ }\bibfield  {title} {\bibinfo {title} {{Planckian dissipation,
  minimal viscosity and the transport in cuprate strange metals}},\ }\href
  {https://doi.org/10.21468/SciPostPhys.6.5.061} {\bibfield  {journal}
  {\bibinfo  {journal} {SciPost Phys.}\ }\textbf {\bibinfo {volume} {6}},\
  \bibinfo {pages} {61} (\bibinfo {year} {2019})}\BibitemShut {NoStop}%
\bibitem [{\citenamefont {Hartnoll}\ and\ \citenamefont
  {Mackenzie}(2021)}]{hartnoll-review}%
  \BibitemOpen
  \bibfield  {author} {\bibinfo {author} {\bibfnamefont {S.~A.}\ \bibnamefont
  {Hartnoll}}\ and\ \bibinfo {author} {\bibfnamefont {A.~P.}\ \bibnamefont
  {Mackenzie}},\ }\href@noop {} {\bibinfo {title} {Planckian dissipation in
  metals}} (\bibinfo {year} {2021}),\ \Eprint
  {https://arxiv.org/abs/2107.07802} {arXiv:2107.07802 [cond-mat.str-el]}
  \BibitemShut {NoStop}%
\bibitem [{\citenamefont {Kovtun}\ \emph {et~al.}(2005)\citenamefont {Kovtun},
  \citenamefont {Son},\ and\ \citenamefont {Starinets}}]{kss}%
  \BibitemOpen
  \bibfield  {author} {\bibinfo {author} {\bibfnamefont {P.~K.}\ \bibnamefont
  {Kovtun}}, \bibinfo {author} {\bibfnamefont {D.~T.}\ \bibnamefont {Son}},\
  and\ \bibinfo {author} {\bibfnamefont {A.~O.}\ \bibnamefont {Starinets}},\
  }\bibfield  {title} {\bibinfo {title} {Viscosity in strongly interacting
  quantum field theories from black hole physics},\ }\href
  {https://doi.org/10.1103/PhysRevLett.94.111601} {\bibfield  {journal}
  {\bibinfo  {journal} {Phys. Rev. Lett.}\ }\textbf {\bibinfo {volume} {94}},\
  \bibinfo {pages} {111601} (\bibinfo {year} {2005})}\BibitemShut {NoStop}%
\bibitem [{\citenamefont {Hartnoll}(2015)}]{Hartnoll2015}%
  \BibitemOpen
  \bibfield  {author} {\bibinfo {author} {\bibfnamefont {S.~A.}\ \bibnamefont
  {Hartnoll}},\ }\bibfield  {title} {\bibinfo {title} {Theory of universal
  incoherent metallic transport},\ }\href {https://doi.org/10.1038/nphys3174}
  {\bibfield  {journal} {\bibinfo  {journal} {Nature Physics}\ }\textbf
  {\bibinfo {volume} {11}},\ \bibinfo {pages} {54} (\bibinfo {year}
  {2015})}\BibitemShut {NoStop}%
\bibitem [{\citenamefont {Allen}\ and\ \citenamefont
  {Mitrović}(1983)}]{ALLEN19831}%
  \BibitemOpen
  \bibfield  {author} {\bibinfo {author} {\bibfnamefont {P.~B.}\ \bibnamefont
  {Allen}}\ and\ \bibinfo {author} {\bibfnamefont {B.}~\bibnamefont
  {Mitrović}},\ }\bibfield  {title} {\bibinfo {title} {Theory of
  superconducting {T}c}\ }(\bibinfo  {publisher} {Academic Press},\ \bibinfo
  {year} {1983})\ pp.\ \bibinfo {pages} {1--92}\BibitemShut {NoStop}%
\bibitem [{\citenamefont {Grimvall}(1976)}]{Grimvall_1976}%
  \BibitemOpen
  \bibfield  {author} {\bibinfo {author} {\bibfnamefont {G.}~\bibnamefont
  {Grimvall}},\ }\bibfield  {title} {\bibinfo {title} {The electron-phonon
  interaction in normal metals},\ }\href
  {https://doi.org/10.1088/0031-8949/14/1-2/013} {\bibfield  {journal}
  {\bibinfo  {journal} {Physica Scripta}\ }\textbf {\bibinfo {volume} {14}},\
  \bibinfo {pages} {63} (\bibinfo {year} {1976})}\BibitemShut {NoStop}%
\bibitem [{\citenamefont {Allen}(1987)}]{allen-lambdatr}%
  \BibitemOpen
  \bibfield  {author} {\bibinfo {author} {\bibfnamefont {P.~B.}\ \bibnamefont
  {Allen}},\ }\bibfield  {title} {\bibinfo {title} {Empirical electron-phonon
  $\ensuremath{\lambda}$ values from resistivity of cubic metallic elements},\
  }\href {https://doi.org/10.1103/PhysRevB.36.2920} {\bibfield  {journal}
  {\bibinfo  {journal} {Phys. Rev. B}\ }\textbf {\bibinfo {volume} {36}},\
  \bibinfo {pages} {2920} (\bibinfo {year} {1987})}\BibitemShut {NoStop}%
\bibitem [{\citenamefont {Wu}\ \emph {et~al.}(2019)\citenamefont {Wu},
  \citenamefont {Hwang},\ and\ \citenamefont {Das~Sarma}}]{sds-magic}%
  \BibitemOpen
  \bibfield  {author} {\bibinfo {author} {\bibfnamefont {F.}~\bibnamefont
  {Wu}}, \bibinfo {author} {\bibfnamefont {E.}~\bibnamefont {Hwang}},\ and\
  \bibinfo {author} {\bibfnamefont {S.}~\bibnamefont {Das~Sarma}},\ }\bibfield
  {title} {\bibinfo {title} {Phonon-induced giant linear-in-{$T$} resistivity
  in magic angle twisted bilayer graphene: Ordinary strangeness and exotic
  superconductivity},\ }\href {https://doi.org/10.1103/PhysRevB.99.165112}
  {\bibfield  {journal} {\bibinfo  {journal} {Phys. Rev. B}\ }\textbf {\bibinfo
  {volume} {99}},\ \bibinfo {pages} {165112} (\bibinfo {year}
  {2019})}\BibitemShut {NoStop}%
\bibitem [{\citenamefont {Sadovskii}(2021)}]{Sadovskii_2021}%
  \BibitemOpen
  \bibfield  {author} {\bibinfo {author} {\bibfnamefont {M.~V.}\ \bibnamefont
  {Sadovskii}},\ }\bibfield  {title} {\bibinfo {title} {Planckian relaxation
  delusion in metals},\ }\href {https://doi.org/10.3367/ufne.2020.08.038821}
  {\bibfield  {journal} {\bibinfo  {journal} {Physics-Uspekhi}\ }\textbf
  {\bibinfo {volume} {64}},\ \bibinfo {pages} {175} (\bibinfo {year}
  {2021})}\BibitemShut {NoStop}%
\bibitem [{\citenamefont {Hwang}\ and\ \citenamefont
  {Das~Sarma}(2019)}]{sds-dilute}%
  \BibitemOpen
  \bibfield  {author} {\bibinfo {author} {\bibfnamefont {E.~H.}\ \bibnamefont
  {Hwang}}\ and\ \bibinfo {author} {\bibfnamefont {S.}~\bibnamefont
  {Das~Sarma}},\ }\bibfield  {title} {\bibinfo {title} {Linear-in-{$T$}
  resistivity in dilute metals: A {F}ermi liquid perspective},\ }\href
  {https://doi.org/10.1103/PhysRevB.99.085105} {\bibfield  {journal} {\bibinfo
  {journal} {Phys. Rev. B}\ }\textbf {\bibinfo {volume} {99}},\ \bibinfo
  {pages} {085105} (\bibinfo {year} {2019})}\BibitemShut {NoStop}%
\bibitem [{\citenamefont {Cao}\ \emph {et~al.}(2020)\citenamefont {Cao},
  \citenamefont {Chowdhury}, \citenamefont {Rodan-Legrain}, \citenamefont
  {Rubies-Bigorda}, \citenamefont {Watanabe}, \citenamefont {Taniguchi},
  \citenamefont {Senthil},\ and\ \citenamefont
  {Jarillo-Herrero}}]{mit-planckian}%
  \BibitemOpen
  \bibfield  {author} {\bibinfo {author} {\bibfnamefont {Y.}~\bibnamefont
  {Cao}}, \bibinfo {author} {\bibfnamefont {D.}~\bibnamefont {Chowdhury}},
  \bibinfo {author} {\bibfnamefont {D.}~\bibnamefont {Rodan-Legrain}}, \bibinfo
  {author} {\bibfnamefont {O.}~\bibnamefont {Rubies-Bigorda}}, \bibinfo
  {author} {\bibfnamefont {K.}~\bibnamefont {Watanabe}}, \bibinfo {author}
  {\bibfnamefont {T.}~\bibnamefont {Taniguchi}}, \bibinfo {author}
  {\bibfnamefont {T.}~\bibnamefont {Senthil}},\ and\ \bibinfo {author}
  {\bibfnamefont {P.}~\bibnamefont {Jarillo-Herrero}},\ }\bibfield  {title}
  {\bibinfo {title} {Strange metal in magic-angle graphene with near
  {P}lanckian dissipation},\ }\href
  {https://doi.org/10.1103/PhysRevLett.124.076801} {\bibfield  {journal}
  {\bibinfo  {journal} {Phys. Rev. Lett.}\ }\textbf {\bibinfo {volume} {124}},\
  \bibinfo {pages} {076801} (\bibinfo {year} {2020})}\BibitemShut {NoStop}%
\bibitem [{\citenamefont {Polshyn}\ \emph {et~al.}(2019)\citenamefont
  {Polshyn}, \citenamefont {Yankowitz}, \citenamefont {Chen}, \citenamefont
  {Zhang}, \citenamefont {Watanabe}, \citenamefont {Taniguchi}, \citenamefont
  {Dean},\ and\ \citenamefont {Young}}]{graphene-linear}%
  \BibitemOpen
  \bibfield  {author} {\bibinfo {author} {\bibfnamefont {H.}~\bibnamefont
  {Polshyn}}, \bibinfo {author} {\bibfnamefont {M.}~\bibnamefont {Yankowitz}},
  \bibinfo {author} {\bibfnamefont {S.}~\bibnamefont {Chen}}, \bibinfo {author}
  {\bibfnamefont {Y.}~\bibnamefont {Zhang}}, \bibinfo {author} {\bibfnamefont
  {K.}~\bibnamefont {Watanabe}}, \bibinfo {author} {\bibfnamefont
  {T.}~\bibnamefont {Taniguchi}}, \bibinfo {author} {\bibfnamefont {C.~R.}\
  \bibnamefont {Dean}},\ and\ \bibinfo {author} {\bibfnamefont {A.~F.}\
  \bibnamefont {Young}},\ }\bibfield  {title} {\bibinfo {title} {Large
  linear-in-temperature resistivity in twisted bilayer graphene},\ }\href
  {https://doi.org/10.1038/s41567-019-0596-3} {\bibfield  {journal} {\bibinfo
  {journal} {Nature Physics}\ }\textbf {\bibinfo {volume} {15}},\ \bibinfo
  {pages} {1011} (\bibinfo {year} {2019})}\BibitemShut {NoStop}%
\bibitem [{\citenamefont {Buterakos}\ and\ \citenamefont
  {Das~Sarma}(2019)}]{sds-trivial-nfl}%
  \BibitemOpen
  \bibfield  {author} {\bibinfo {author} {\bibfnamefont {D.}~\bibnamefont
  {Buterakos}}\ and\ \bibinfo {author} {\bibfnamefont {S.}~\bibnamefont
  {Das~Sarma}},\ }\bibfield  {title} {\bibinfo {title} {Coupled
  electron-impurity and electron-phonon systems as trivial non-{F}ermi
  liquids},\ }\href {https://doi.org/10.1103/PhysRevB.100.235149} {\bibfield
  {journal} {\bibinfo  {journal} {Phys. Rev. B}\ }\textbf {\bibinfo {volume}
  {100}},\ \bibinfo {pages} {235149} (\bibinfo {year} {2019})}\BibitemShut
  {NoStop}%
\bibitem [{\citenamefont {Armitage}\ \emph {et~al.}(2010)\citenamefont
  {Armitage}, \citenamefont {Fournier},\ and\ \citenamefont
  {Greene}}]{rick-rmp}%
  \BibitemOpen
  \bibfield  {author} {\bibinfo {author} {\bibfnamefont {N.~P.}\ \bibnamefont
  {Armitage}}, \bibinfo {author} {\bibfnamefont {P.}~\bibnamefont {Fournier}},\
  and\ \bibinfo {author} {\bibfnamefont {R.~L.}\ \bibnamefont {Greene}},\
  }\bibfield  {title} {\bibinfo {title} {Progress and perspectives on
  electron-doped cuprates},\ }\href
  {https://doi.org/10.1103/RevModPhys.82.2421} {\bibfield  {journal} {\bibinfo
  {journal} {Rev. Mod. Phys.}\ }\textbf {\bibinfo {volume} {82}},\ \bibinfo
  {pages} {2421} (\bibinfo {year} {2010})}\BibitemShut {NoStop}%
\bibitem [{\citenamefont {Sarkar}\ \emph {et~al.}(2017)\citenamefont {Sarkar},
  \citenamefont {Mandal}, \citenamefont {Higgins}, \citenamefont {Zhao},
  \citenamefont {Yu}, \citenamefont {Jin},\ and\ \citenamefont
  {Greene}}]{tara-prb}%
  \BibitemOpen
  \bibfield  {author} {\bibinfo {author} {\bibfnamefont {T.}~\bibnamefont
  {Sarkar}}, \bibinfo {author} {\bibfnamefont {P.~R.}\ \bibnamefont {Mandal}},
  \bibinfo {author} {\bibfnamefont {J.~S.}\ \bibnamefont {Higgins}}, \bibinfo
  {author} {\bibfnamefont {Y.}~\bibnamefont {Zhao}}, \bibinfo {author}
  {\bibfnamefont {H.}~\bibnamefont {Yu}}, \bibinfo {author} {\bibfnamefont
  {K.}~\bibnamefont {Jin}},\ and\ \bibinfo {author} {\bibfnamefont {R.~L.}\
  \bibnamefont {Greene}},\ }\bibfield  {title} {\bibinfo {title} {Fermi surface
  reconstruction and anomalous low-temperature resistivity in electron-doped
  $\mathrm{L}{\mathrm{a}}_{2\ensuremath{-}x}\mathrm{C}{\mathrm{e}}_{x}\mathrm{Cu}{\mathrm{o}}_{4}$},\
  }\href {https://doi.org/10.1103/PhysRevB.96.155449} {\bibfield  {journal}
  {\bibinfo  {journal} {Phys. Rev. B}\ }\textbf {\bibinfo {volume} {96}},\
  \bibinfo {pages} {155449} (\bibinfo {year} {2017})}\BibitemShut {NoStop}%
\bibitem [{\citenamefont {Dagan}\ \emph {et~al.}(2004)\citenamefont {Dagan},
  \citenamefont {Qazilbash}, \citenamefont {Hill}, \citenamefont {Kulkarni},\
  and\ \citenamefont {Greene}}]{yoram-prl}%
  \BibitemOpen
  \bibfield  {author} {\bibinfo {author} {\bibfnamefont {Y.}~\bibnamefont
  {Dagan}}, \bibinfo {author} {\bibfnamefont {M.~M.}\ \bibnamefont
  {Qazilbash}}, \bibinfo {author} {\bibfnamefont {C.~P.}\ \bibnamefont {Hill}},
  \bibinfo {author} {\bibfnamefont {V.~N.}\ \bibnamefont {Kulkarni}},\ and\
  \bibinfo {author} {\bibfnamefont {R.~L.}\ \bibnamefont {Greene}},\ }\bibfield
   {title} {\bibinfo {title} {Evidence for a quantum phase transition in
  {${\mathrm{Pr}}_{2\ensuremath{-}x}{\mathrm{Ce}}_{x}{\mathrm{CuO}}_{4\ensuremath{-}\ensuremath{\delta}}$}
  from transport measurements},\ }\href
  {https://doi.org/10.1103/PhysRevLett.92.167001} {\bibfield  {journal}
  {\bibinfo  {journal} {Phys. Rev. Lett.}\ }\textbf {\bibinfo {volume} {92}},\
  \bibinfo {pages} {167001} (\bibinfo {year} {2004})}\BibitemShut {NoStop}%
\bibitem [{\citenamefont {Jin}\ \emph {et~al.}(2011)\citenamefont {Jin},
  \citenamefont {Butch}, \citenamefont {Kirshenbaum}, \citenamefont
  {Paglione},\ and\ \citenamefont {Greene}}]{kui}%
  \BibitemOpen
  \bibfield  {author} {\bibinfo {author} {\bibfnamefont {K.}~\bibnamefont
  {Jin}}, \bibinfo {author} {\bibfnamefont {N.~P.}\ \bibnamefont {Butch}},
  \bibinfo {author} {\bibfnamefont {K.}~\bibnamefont {Kirshenbaum}}, \bibinfo
  {author} {\bibfnamefont {J.}~\bibnamefont {Paglione}},\ and\ \bibinfo
  {author} {\bibfnamefont {R.~L.}\ \bibnamefont {Greene}},\ }\bibfield  {title}
  {\bibinfo {title} {Link between spin fluctuations and electron pairing in
  copper oxide superconductors},\ }\href {https://doi.org/10.1038/nature10308}
  {\bibfield  {journal} {\bibinfo  {journal} {Nature}\ }\textbf {\bibinfo
  {volume} {476}},\ \bibinfo {pages} {73} (\bibinfo {year} {2011})}\BibitemShut
  {NoStop}%
\bibitem [{\citenamefont {Fournier}\ \emph {et~al.}(1998)\citenamefont
  {Fournier}, \citenamefont {Mohanty}, \citenamefont {Maiser}, \citenamefont
  {Darzens}, \citenamefont {Venkatesan}, \citenamefont {Lobb}, \citenamefont
  {Czjzek}, \citenamefont {Webb},\ and\ \citenamefont
  {Greene}}]{pcco-linear-t}%
  \BibitemOpen
  \bibfield  {author} {\bibinfo {author} {\bibfnamefont {P.}~\bibnamefont
  {Fournier}}, \bibinfo {author} {\bibfnamefont {P.}~\bibnamefont {Mohanty}},
  \bibinfo {author} {\bibfnamefont {E.}~\bibnamefont {Maiser}}, \bibinfo
  {author} {\bibfnamefont {S.}~\bibnamefont {Darzens}}, \bibinfo {author}
  {\bibfnamefont {T.}~\bibnamefont {Venkatesan}}, \bibinfo {author}
  {\bibfnamefont {C.~J.}\ \bibnamefont {Lobb}}, \bibinfo {author}
  {\bibfnamefont {G.}~\bibnamefont {Czjzek}}, \bibinfo {author} {\bibfnamefont
  {R.~A.}\ \bibnamefont {Webb}},\ and\ \bibinfo {author} {\bibfnamefont
  {R.~L.}\ \bibnamefont {Greene}},\ }\bibfield  {title} {\bibinfo {title}
  {Insulator-metal crossover near optimal doping in
  {${\mathrm{Pr}}_{2\ensuremath{-}\mathit{x}}{\mathrm{Ce}}_{\mathit{x}}{\mathrm{CuO}}_{4}$}:
  Anomalous normal-state low temperature resistivity},\ }\href
  {https://doi.org/10.1103/PhysRevLett.81.4720} {\bibfield  {journal} {\bibinfo
   {journal} {Phys. Rev. Lett.}\ }\textbf {\bibinfo {volume} {81}},\ \bibinfo
  {pages} {4720} (\bibinfo {year} {1998})}\BibitemShut {NoStop}%
\bibitem [{\citenamefont {Sarkar}\ \emph {et~al.}(2018)\citenamefont {Sarkar},
  \citenamefont {Greene},\ and\ \citenamefont {Das~Sarma}}]{fl-sm}%
  \BibitemOpen
  \bibfield  {author} {\bibinfo {author} {\bibfnamefont {T.}~\bibnamefont
  {Sarkar}}, \bibinfo {author} {\bibfnamefont {R.~L.}\ \bibnamefont {Greene}},\
  and\ \bibinfo {author} {\bibfnamefont {S.}~\bibnamefont {Das~Sarma}},\
  }\bibfield  {title} {\bibinfo {title} {Anomalous normal-state resistivity in
  superconducting
  $\mathrm{L}{\mathrm{a}}_{2\ensuremath{-}x}\mathrm{C}{\mathrm{e}}_{x}\mathrm{Cu}{\mathrm{o}}_{4}$:
  Fermi liquid or strange metal?},\ }\href
  {https://doi.org/10.1103/PhysRevB.98.224503} {\bibfield  {journal} {\bibinfo
  {journal} {Phys. Rev. B}\ }\textbf {\bibinfo {volume} {98}},\ \bibinfo
  {pages} {224503} (\bibinfo {year} {2018})}\BibitemShut {NoStop}%
\bibitem [{\citenamefont {Zimmers}\ \emph {et~al.}(2005)\citenamefont
  {Zimmers}, \citenamefont {Tomczak}, \citenamefont {Lobo}, \citenamefont
  {Bontemps}, \citenamefont {Hill}, \citenamefont {Barr}, \citenamefont
  {Dagan}, \citenamefont {Greene}, \citenamefont {Millis},\ and\ \citenamefont
  {Homes}}]{Zimmers_2005}%
  \BibitemOpen
  \bibfield  {author} {\bibinfo {author} {\bibfnamefont {A.}~\bibnamefont
  {Zimmers}}, \bibinfo {author} {\bibfnamefont {J.~M.}\ \bibnamefont
  {Tomczak}}, \bibinfo {author} {\bibfnamefont {R.~P. S.~M.}\ \bibnamefont
  {Lobo}}, \bibinfo {author} {\bibfnamefont {N.}~\bibnamefont {Bontemps}},
  \bibinfo {author} {\bibfnamefont {C.~P.}\ \bibnamefont {Hill}}, \bibinfo
  {author} {\bibfnamefont {M.~C.}\ \bibnamefont {Barr}}, \bibinfo {author}
  {\bibfnamefont {Y.}~\bibnamefont {Dagan}}, \bibinfo {author} {\bibfnamefont
  {R.~L.}\ \bibnamefont {Greene}}, \bibinfo {author} {\bibfnamefont {A.~J.}\
  \bibnamefont {Millis}},\ and\ \bibinfo {author} {\bibfnamefont {C.~C.}\
  \bibnamefont {Homes}},\ }\bibfield  {title} {\bibinfo {title} {Infrared
  properties of electron-doped cuprates: Tracking normal-state gaps and quantum
  critical behavior in {Pr$_{2-x}$Ce$_{x}$CuO$_4$}},\ }\href
  {https://doi.org/10.1209/epl/i2004-10480-2} {\bibfield  {journal} {\bibinfo
  {journal} {Europhysics Letters ({EPL})}\ }\textbf {\bibinfo {volume} {70}},\
  \bibinfo {pages} {225} (\bibinfo {year} {2005})}\BibitemShut {NoStop}%
\bibitem [{\citenamefont {Collignon}\ \emph {et~al.}(2020)\citenamefont
  {Collignon}, \citenamefont {Bourges}, \citenamefont {Fauqu\'e},\ and\
  \citenamefont {Behnia}}]{sto-prx}%
  \BibitemOpen
  \bibfield  {author} {\bibinfo {author} {\bibfnamefont {C.}~\bibnamefont
  {Collignon}}, \bibinfo {author} {\bibfnamefont {P.}~\bibnamefont {Bourges}},
  \bibinfo {author} {\bibfnamefont {B.}~\bibnamefont {Fauqu\'e}},\ and\
  \bibinfo {author} {\bibfnamefont {K.}~\bibnamefont {Behnia}},\ }\bibfield
  {title} {\bibinfo {title} {Heavy nondegenerate electrons in doped strontium
  titanate},\ }\href {https://doi.org/10.1103/PhysRevX.10.031025} {\bibfield
  {journal} {\bibinfo  {journal} {Phys. Rev. X}\ }\textbf {\bibinfo {volume}
  {10}},\ \bibinfo {pages} {031025} (\bibinfo {year} {2020})}\BibitemShut
  {NoStop}%
\end{thebibliography}%

\end{document}